\documentclass[aps,prb,reprint,a4paper]{revtex4-1}
\usepackage[latin1]{inputenc}
\usepackage{amsmath}
\usepackage{amsfonts}
\usepackage{amssymb}
\usepackage{graphicx}
\usepackage{color}
\newcommand{\SD}{{\text{SD}}}

\begin{document}

\title
{Salient Signature of van der Waals Interactions}
\author{Mireia Via-Nadal,$^1$ Mauricio Rodr\'iguez-Mayorga,$^{1,2}$ and Eduard Matito$^{1,3}$}
   \email[To whom the correspondence should be addressed: ]{ematito@gmail.com}
   \affiliation{$^1$Kimika Fakultatea, Euskal Herriko Unibertsitatea (UPV/EHU), and Donostia
International Physics Center (DIPC), P.K. 1072, 20080 Donostia, Euskadi, Spain.}
   \affiliation{$^2$Institut de Qu\'imica Computacional i Cat\`alisi (IQCC) and Departament de Qu\'imica, University of Girona, 17071 Girona, Catalonia, Spain,}
   \affiliation{$^3$IKERBASQUE, Basque Foundation for Science, 48011 Bilbao, Euskadi, Spain.}
\date{\today}

\begin{abstract}
van der Waals interactions govern the physics of a plethora of molecular structures.
It is well known that the leading term in the distance-based London expansion of the 
van der Waals
energy for atomic and molecular dimers decays as $1/R^6$, where $R$ is the dimer distance.
Using perturbation theory, we find the leading term in the distance-based expansion
of the intracule pair density at the interatomic distance.
Our results unveil a universal $1/R^3$ decay, which is less prone to numerical errors than the
$1/R^6$ dependency, and it is confirmed numerically in H$_2$ and He$_2$ molecules.
This \textit{signature} of van der Waals interactions can be directly used in the 
construction of approximate pair density and energy functionals including vdW corrections.
\end{abstract}

\maketitle


Dispersion or van der Waals (vdW) interactions are ubiquitous in nature, governing the 
stability of molecules and materials,\cite{hermann:17cr}
and having an essential role in molecular recognition,\cite{yang:13natc}
the double-helical structure of DNA,\cite{cerny:08jacs}
molecular adsorption on surfaces,\cite{klimes:11prb,rosa:14prb}
and the adhesion of micromachined surfaces.\cite{delrio:05natm}
They are so important in physics, chemistry and biology, that even the most simple
electronic structure methods consider corrections for vdW interactions.
Due to their long-range dynamic-correlation nature, they are not well modelled by
standard functionals in density functional theory (DFT),\cite{hermann:17cr}
which by construction are essentially local or semi-local in nature.~\cite{kristyan:94cpl}
Hence, except for a few functionals,\cite{berland:15rpp,thonhauser:07prb}
most DFT functionals include \textit{ad hoc} energy corrections to 
account for vdW interactions.\cite{grimme:11wires,becke:05jcp,becke:07jcp}

vdW forces arise from the electrostatic interaction between fluctuations in the electron density,
and the pairwise effect in the energy shows a leading $1/R^6$ dependency, where $R$ is the interaction
distance between fragments. 
This fact is often exploited in the construction of effective pairwise potentials that enter
the expressions of various methods. On the other hand, the effect of vdW interactions in the
wavefunction or related quantities has been less discussed in the 
literature.~\cite{rapcewicz:91prb,hyldgaard:14prb,ferri:15prl,ferri:17prm}
This knowledge could shed some light in the design of computational approaches including
vdW interactions and provide further tests to calibrate electronic structure theory
methods.

In this letter we use perturbation theory to find the leading term in the expansion
of the intracule pair density in terms of $R$, the interatomic distance.
Our results reveal a universal $1/R^3$ 
dependency that is corroborated by numerical calculations in H$_2$ and He$_2$ molecules. 
Upon integration of the vdW contribution to the intracule we recover the
vdW energy that follows the stablished $1/R^6$ dependency.

\section{Theory}

\begin{figure}
\begin{center}
\includegraphics[scale=0.07]{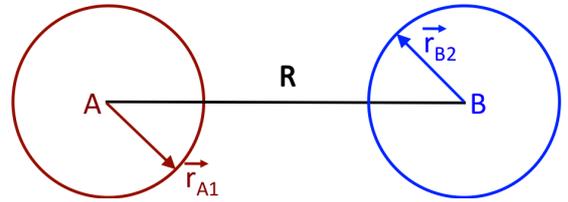}
\caption{Two atoms, $A$ and $B$, separated by a distance $R$ and two electrons with
coordinates $\vec{r}_{A1}$ and $\vec{r}_{B2}$ defined with respect to the position of atoms
$A$ and $B$.}
\label{frame}
\end{center}
\end{figure}

We start from the unperturbed wavefunction for a system of two hydrogen atoms, $A$ and $B$, 
separated a distance $R$, given by the product of two hydrogenoid $1s$ functions,\cite{pauling:35book}
\begin{equation}\label{eq:psi0}
\Psi^{(0)}(\mathbf{r}_{1},\mathbf{r}_{2}) = N\phi{}_{A}(\mathbf{r}_{1})\phi{}_{B}(\mathbf{r}_{2}) \,.
\end{equation}
For convenience we have chosen to represent the hydrogenoid functions 
by one Gaussian function of exponent $\alpha$. This change does not alter the $1/R^6$
dependency on the calculation of the second-order correction to the energy (\textit{vide infra}).

The perturbation operator contains all the interactions between fragments $A$ and $B$:
\begin{equation}\label{eq:H1}
\hat{H}^{(1)}=-\frac{1}{r_{A2}}-\frac{1}{r_{B1}}+\frac{1}{R}+\frac{1}{r_{12}} \,,
\end{equation}
where we have labeled the electrons in $A$ and $B$ as 1 and 2, respectively (see Fig.~\ref{frame}).
After application of Uns\"old's approximation,\cite{unsold:27zp,hirschfelder:64aqc}
the expansion of Eq.~\ref{eq:H1} in terms of $1/R$ at large distances
yields the first-order correction to the wavefunction,
\begin{eqnarray}\nonumber
\Psi^{(1)}(\mathbf{r}_{1},\mathbf{r}_{2}) &\approx& 
\frac{\Psi^{(0)}(\mathbf{r}_{1},\mathbf{r}_{2})}{R^3}\left(x_{A1}x_{B2}+y_{A1}y_{B2}-2z_{A1}z_{B2}\right) \\ &+& \mathcal{O}\left(R^{-4}\right) \,,
\end{eqnarray}
where we have assumed that the molecule is located in the z-axis and $x_{A1},y_{B2}$, etc.
refer to the Cartesian components of the vectors in Eq.~\ref{eq:H1}.
Using the first-order correction to the wavefunction we can demonstrate that the
leading term in the second-order correction to the energy follows the widely known
$1/R^6$ dependency\cite{pauling:35book} 
\begin{equation}
E^{(2)}=\frac{6}{\alpha^4 R^6}+\mathcal{O}\left(R^{-8}\right) \,.
\end{equation}

Now, let us consider the first-order correction to the pair density, 
\begin{eqnarray}
n_{2}^{(1)}(\mathbf{r}_{1},\mathbf{\mathbf{r}}_{2})  = 
2\Psi^{(0)}(\mathbf{r}_{1},\mathbf{\mathbf{r}}_{2})\Psi^{(1)}(\mathbf{r}_{1},\mathbf{\mathbf{r}}_{2}) \,.
\end{eqnarray}
The intracule of the pair density is a function that provides a distribution of
the interelectronic separations,\cite{coulson:61ppsl}
\begin{equation}
I(u)=\int\int d\mathbf{r}_{1}d\mathbf{r}_{2} \,
n_2(\mathbf{r}_{1},\mathbf{r}_{2})\delta(u-r_{12})  \,,
\end{equation}
and thus returns the average electron-electron distance upon integration over
$u$.
The intracule of the pair density is actually connected with an experimental observable, 
the X-ray scattering intensity, which is essentially determined by the Fourier-Bessel transform
of the radial intracule probability density.~\cite{thakkar:84ijqc,thakkar:84pra}

After some algebraic manipulation one can prove that the zeroth-order intracule of
the pair density at $R$,
\begin{equation}\label{eq:i0}
I^{(0)}(R)=\left(\frac{\alpha}{16 \pi^3}\right)^{1/2}\left(1-e^{-4\alpha R^2}\right) \,,
\end{equation}
yields a constant value in the limit, which corresponds to the distribution
of two independent electrons. 
A Gaussian function enters the expression in Eq.~\ref{eq:i0} and, therefore,
the form of the zeroth-order intracule function at $R$ depends on the reference wavefunction,
Eq.~\ref{eq:psi0}.
Conversely, the first-order correction at $R$ decays as $1/R^3$, 
\begin{equation}\label{r3}
\lim_{R\to\infty}I^{(1)}(R)=-\frac{4\left(1+8\sqrt{2}\right)\alpha^{5/2}}{\pi^{7/2}R^{3}} \,,
\end{equation}
without any exponential terms in the expression,
suggesting that the $1/R^3$ dependency does not rely on a particular choice of 
the zeroth-order reference.
As we will check numerically in the next section, 
Eq.~\ref{r3} puts forward
a universal condition that can be employed to assess the performance of approximate
pair densities and models of the intracule function in reproducing vdW interactions.

\section{Numerical Examples}

As a zeroth-order pair density we choose 
\begin{equation}\label{eq:2pd0}
n_2^{\SD}(\mathbf{r}_1,\mathbf{r}_2)=n(\mathbf{r}_1)n(\mathbf{r}_2)
- n_1(\mathbf{r}_1;\mathbf{r}_2)n_1(\mathbf{r}_2;\mathbf{r}_1)   \,, 
\end{equation}
\vspace{0.2cm}
where $n_1(\mathbf{r}_1;\mathbf{r}_2)$ is the first-order reduced density matrix.
$n_2^{\SD}$ is the minimal model that guarantees that 
$n_2^{\SD}(\mathbf{r}_1,\mathbf{r}_2)\rightarrow n(\mathbf{r}_1)n(\mathbf{r}_2)$ 
at large interatomic distances
and, at the same time, preserves the antisymmetric nature of particles,\cite{rodriguez:17pccp2}
which is necessary to remove the spin entanglement effects that also appear at large
interelectronic separations and are not included by the second-order perturbational treatment.
Hence, we will evaluate the intracule resulting from the following pair density
difference
\begin{equation}\label{eq:i2}
I^{(1)}(u)\approx\int\int
d\mathbf{r}_{1}d\mathbf{r}_{2}
\left[n_2(\mathbf{r}_1,\mathbf{r}_2)-n_2^{\SD}(\mathbf{r}_1,\mathbf{r}_2)\right]
\delta(u-r_{12})
\end{equation}

\begin{figure}[h]
\centering
\includegraphics[scale=0.76]{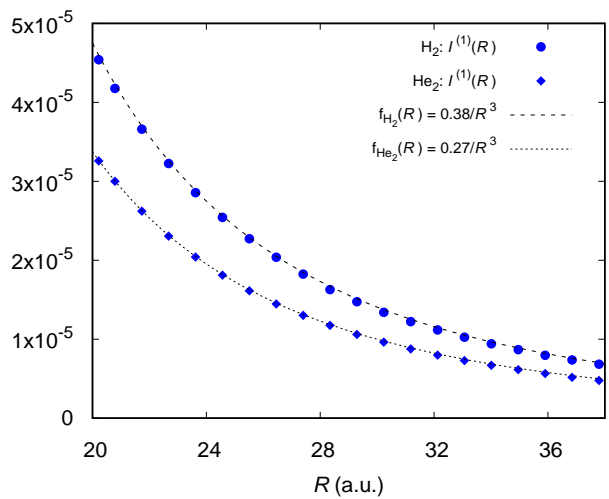}
\caption{$I^{(1)}(R)$ against $R$, the interatomic distance, for H$_2$ and He$_2$.
The fitting corresponds to $f(R)= a / R^{3}$, where $a$ is the fitted parameter and equals
0.38 and 0.27 for H$_2$ and He$_2$, respectively. In both cases the Pearson regression 
coefficient, $r^2$, is greater than 0.999. All quantities are in a.u.}
\label{fig:intra}
\end{figure}

To this aim, we have chosen two simple molecules, H$_2$ and He$_2$, and we have performed
full-configuration interaction calculations~\footnote{The calculations have been performed
with a modified version of the program of Knowles and Handy~\cite{knowles:89cpc} and the
pair density matrices have been obtained with our in-house DMN code.~\cite{dmn}
The intracule calculation employed the in-house RHO2\_OPS\cite{rho2ops} code, which uses the 
algorithm\cite{cioslowski:96jcpfast} of Cioslowski and Liu.}
with the aug-cc-pVDZ basis set at different interatomic separations ($R$).
We have computed Eq.~\ref{eq:i2} and plotted $I^{(1)}(R)$ against $R$ (see Fig.~\ref{fig:intra}).

The numerical results confirm the predicted $1/R^{3}$ dependency of the vdW correction
to the intracule of the pair density. The fitting procedure employs the points presented
in Fig.~\ref{fig:intra} and uses a least squares minimization analysis to determine the
Pearson regression coefficient and the $a$ parameter in the $f(R)= a / R^{3}$ fitting function.\newline

The intracule pair density is the simplest quantity in terms of which we can express explicitly
and exactly the electron-electron repulsion energy, 
\begin{equation}
V_{ee}=\int du \frac{I(u)}{u}   \,,
\end{equation}
and the integration of the first-order correction to the intracule divided by the 
electron-electron distance provides the vdW energy:
\begin{equation}\label{eq:vee2}
V_{ee}^{(2)}=\int du \frac{I^{(1)}(u)}{u}  \,.
\end{equation}

\begin{figure}[h]
\centering
\includegraphics[scale=0.76]{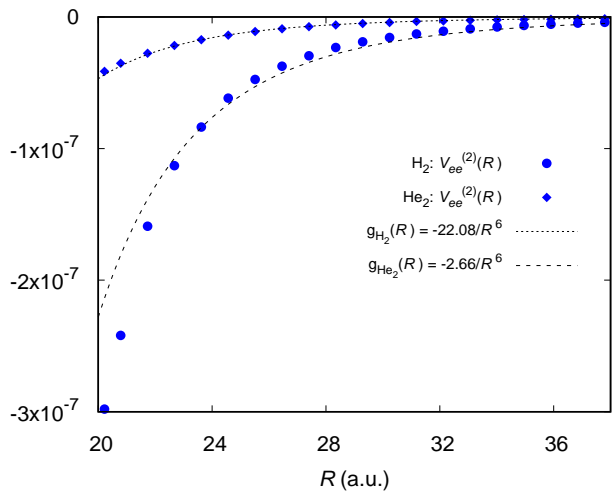}
\caption{$V_{ee}^{(2)}(R)$ against $R$, the interatomic distance, for H$_2$ and He$_2$.
The fitting corresponds to $g(R)= a / R^{6}$, where $a$ is the fitted parameter and equals
-22.08 and -2.66 for H$_2$ and He$_2$, respectively. In both cases the Pearson coefficient, $r^2$,
is greater than 0.97. All quantities are in a.u.}
\label{fig:vee}
\end{figure}

Through Eq.~\ref{eq:vee2}, we can calculate the vdW energy
using the first-order correction to the intracule function.
In Fig.~\ref{fig:vee} we observe that 
at large distances these functions reproduce the $1/R^{6}$ dependency. 
The fitting is not as good as in the intracule function because of the large power
dependency, which increases the numerical error.

\section{Conclusions}
We have unveiled a universal condition of the intracule of the pair density:
the vdW contribution to the intracule of the pair density at $R$ should decay as
$1/R^3$, $R$ being the separation of two fragments. 
This condition is connected to the well-known
$1/R^6$ decay of the vdW energy and it can be recovered from the vdW contribution
to the intracule of the pair density (see Eq.~\ref{eq:vee2}). 
This requirement is a salient signature of vdW interactions
that can be employed as a stringent constraint in 
a judicious construction of new methods and approximations in electronic structure theory
including vdW interactions.\cite{mazziotti:12prl,schlimgen:16jpcl,mazziotti:16prl} 
Since the vdW correction to the intracule of the pair density 
shows a lower power dependency than the energy one, it
is also less prone to numerical errors (compare fittings in Figs.~\ref{fig:intra} and~\ref{fig:vee}).

\begin{acknowledgements}
The authors thank Dr. Eloy Ramos-Cordoba for helpful discussions.
This research has been funded by Spanish MINECO/FEDER Project CTQ2014-52525-P
and the Basque Country 
Consolidated Group Project No. IT588-13. 
M.V.N. and M.R.M. acknowledge the Spanish 
Ministry of Economy, Industry and Competitiveness (MINECO)
and the Spanish Ministry of Education, Culture and Sports 
for the doctoral grants BES-2015-072734 and FPU-2013/00176, respectively.
\end{acknowledgements}

\bibliography{gen}

\end{document}